# History-dependent dissipative vortex dynamics in superconducting arrays

Malcolm Durkin, Ian Mondragon-Shem, Serena Eley[1], Taylor L. Hughes, Nadya Mason

Department of Physics and Frederick Seitz Materials Research Laboratory, University of Illinois at Urbana-Champaign, Urbana, IL 61801-2902, USA

We perform current($I$)-voltage($V$) measurements on low resistance superconductor-normal-superconductor arrays in finite magnetic fields, focusing on the dilute vortex population regime. We observe significant deviations from predicted behavior, notably the absence of a differential resistance peak near the vortex de-pinning current, and a broad linear $I$-$V$ region with an extrapolated $I$ intercept equal to the de-pinning current. Comparing these results to an overdamped molecular vortex model, we find that this behavior can be explained by the presence of a history dependent dissipative force. This approach has not been considered previously, yet is crucial for obtaining a correct description of the vortex dynamics in superconducting arrays.

## I.     Introduction

Vortex motion dominates the electrical transport properties of two-dimensional (2D) superconductors [1,2]. The type of vortex motion, and thus the dissipative transport response, depends largely on the characteristics of the initial equilibrium vortex phase. In finite magnetic fields, 2D superconductors can exhibit many possible vortex phases, for example crystals due to vortex-vortex interactions or an underlying periodic potential. For weak disorder, such crystalline phases devolve into glass phases and can also melt into liquids [3, 4]. When an external force is applied, de-pinning and bulk vortex motion occur, creating non-equilibrium behavior such as elastic or plastic vortex flow. There has been continuing interest in the nature of such vortex de-pinning, which can depend in complex ways on edge or bulk phases [5], and can demonstrate unique phase transitions, e.g., between Mott insulator and metallic states [6,7]. Superconductor-normal-superconductor (SNS) arrays provide highly tunable platforms for studying such vortex behavior and can be used to access a wide variety of phases. Previously, the dynamic behavior of these arrays has been studied using molecular vortex [8, 9] and resistively and capacitively shunted junction (RCSJ) [10] models. While some predictions of these models, such as constant resistance flux-flow, have been experimentally observed, some crucial aspects of transport–such as the predicted differential resistance peaks–are often absent for reasons that have not been well understood [11,12].

To study the dynamic behavior of vortices, we measure transport across SNS arrays in finite magnetic fields, focusing on the dilute vortex population regime where vortex-vortex interactions are negligible. Here, overdamped molecular vortex [8, 9] and RCSJ array [10] models predict $I$-$V$

---
[1] Present Address: Condensed Matter and Magnet Science Group, Los Alamos National Laboratory, Los Alamos, NM 87545

relationships similar to the single vortex or junction case, with vortices behaving as a damped massless particles in a washboard potential defined by the array geometry. We apply a bias current to supply a driving force, which overcomes the barrier supplied by the washboard potential at a de-pinning current, $I_d$. Above $I_d$, the overdamped tilted washboard models predict $V \propto \sqrt{I^2 - I_d^2}$ behavior in the low temperature limit [13]. Important features of these models include a differential resistance peak near $I_d$ and convergence to a linear flux-flow regime that has an *I*-intercept of $I = 0$ at higher currents. Contrary to these predictions, we observe no differential resistance peak near $I_d$ and our linear *I-V* region has a nonzero extrapolated *I* intercept on the order of $I_d$. This observed behavior leads us to consider a novel phenomenological description of the system based on time delayed dissipative forces. This approach has not been introduced before, yet it is crucial for obtaining a correct description of vortex dynamics, even when interactions are negligible. The absence of this predicted peak is not unique to our arrays [11, 12], and thus the modified dissipation terms discussed in this paper should enhance the understanding of a wide range of vortex systems.

## II.    Experimental Measurements

Our devices consist of triangular arrays of mesoscopic Nb islands on top of 10-nm thick Au films patterned for four-point measurements, as shown in Fig. 1(a). The Nb island height is 125 nm, while the edge-to-edge island spacing varies between 390 and 540 nm depending on the sample; additional fabrication details have been reported previously [14]. From the normal state resistivity of the Au films we extract a mean free path of $\ell \approx 13$ nm, an estimated diffusion constant of $D \approx 95$ cm$^2$/s, and a temperature-dependent coherence length $\xi_N(T) \approx 270$ nm/$\sqrt{T}$, where *T* is in units of K [15]. All Nb islands are 260 nm in diameter, which is approximately ten times the dirty-limit coherence length $\xi_{Nb}^0$. Upon cooling, these arrays exhibit a two-step transition to the superconducting state, as can be seen in the typical resistance, *R*, vs temperature, *T*, curve of Fig. 1. The higher-temperature drop represents the temperature at which the individual Nb islands become superconducting. The lower-temperature drop, which we term $T_c$, is the transition of the entire array to a superconducting state. Previous work [15] has shown that the zero-field transition can be associated with a Berezinski-Kosterless-Thouless transition, i.e., is driven by the binding of thermally induced vortex-antivortex pairs in a 2D superconductor [16,17,18].

As shown in Fig. 1(b), we observe large, periodic magnetoresistance oscillations, which provide strong evidence for vortex-dominated transport in the arrays. In arrays of regularly spaced superconducting islands, the vortex population is determined by the applied magnetic field. The number of vortices per unit cell of the array is given by the frustration parameter, $f = \Phi/\Phi_0$, where $\Phi$ is the flux through a unit cell and $\Phi_0$ is the quantum of flux. The island array forms a periodic potential for vortices, with a barrier between islands that must be overcome for vortices to move. Increasing the magnetic field leads to increased interactions between vortices that effectively reduce the force needed to overcome the barriers between the islands, thus increasing the magnetoresistance [19]. However, at special fillings determined by array geometry the vortex lattice is commensurate with the island array, resulting in a strongly pinned vortex lattice that can be observed as dip in the magnetoresistance [20]. The largest dips in resistance occur at $f$ = 0, 1/4, 1/2, 3/4. Lesser dips are also evident at weakly commensurate frustrations $f$ = 1/8, 1/6, 1/3, 3/8, 2/5, 5/8, and 2/3. The depths of the dips are consistent with the theoretical prediction for the ground-state energies at different values of frustration for triangular arrays [21].

We perform DC current-voltage, or *I-V*, measurements to study the dynamic vortex behavior in the arrays, primarily focusing on the dilute vortex population regime below $f$ = 1/10. The applied current provides a Lorentz force and the resultant vortex motion produces a voltage. Fig. 2(a) shows *I-V* curves and Fig. 2(b) shows d$V$/d$I$ curves as a function of magnetic field for an array of islands spaced 390-nm edge-to-edge, where the current, $I$, is less than the junction critical current, $I_c$, and $T$ = 17 mK. The temperature is much lower than the array transition temperature of 410 mK and we do not observe significant temperature dependence of $I_d$ or $I_c$ in the dilute population regime below 150 mK (see Appendix A); this suggests that measurements are occurring in a low temperature regime not dominated by thermally activated vortices. For low magnetic fields and currents, the vortices are pinned and the system has zero resistance; this is shown schematically in region I of Fig. 1(c). As the current is increased, a transition to a finite resistance state occurs when the Lorentz force overcomes vortex pinning [region II of Fig. 1 (c)]. This transition, which occurs at $I_d$, is a measure of the barriers to vortex motion and can be used to characterize the vortex pinning regime.

At higher currents, vortices move with a terminal velocity in the flux flow regime [region III of Fig. 1(c)]. This flux flow regime is manifested by a linear *I-V* relation and a differential resistance that approaches a fixed value at higher current; this behavior can be seen experimentally in Figs.2(a) and 2(b), respectively. The flux flow differential resistance, $R_{ff}$, is well described by the Bardeen-Stephens model [1], which predicts $R_{ff} \sim 2 f R_n$ for normal state resistivity $R_n$. Fig. 2(b) shows very flat $dV/dI$ at $R_{ff}$ in the low filling regime, indicating that the vortices have reached a terminal velocity over a wide range of currents; Fig. 2(c) shows that $R_{ff}$ is linearly proportional to $f$, as expected.

Contrary to theoretical predictions, we do not observe a differential resistance peak near $I_d$, nor any inflection points in the *I-V* measurements on the approach to flux flow. These are essential features of previous models of the transition from pinned behavior. These features appeared because, in order for the array to transition smoothly from de-pinning to flux flow—i.e, from where $V \sim 0$ and $I$ is between 0 and $I_d$, to flux flow with $V \propto I$ and an $I$ intercept of 0—there had to be an inflection point beyond which $d^2V/dI^2 < 0$. The need for a differential resistance peak can be seen in the predicted *I-V* for an overdamped array shown in Fig. 2(d). The simulated curve initially demonstrates $V = 0$ pinned behavior at low currents, but then rapidly increases in $V$ at $I_d$ as it transitions to $V \propto I$, necessitating a maximum slope near $I_d$.

The absence of this peak in the differential resistance has been previously observed and discussed as a consequence of broadening due to finite temperature [12] and the effect of superposing DC and AC driving currents [22], but these explanations are not convincing for our system. The finite temperature explanation in Rzchowski et al. [12] runs contrary to the analytical expression presented in the same work which predicts that the peak should only be broadened by finite temperature and should not disappear. As shown in Fig. 3(a), our simulations (discussed in section III) show that the peak persists even in the presence of significant thermal fluctuations near $I=0$ (evident as finite resistance flux creep), which indicate a much higher temperature than our experiments. Furthermore, the lack of temperature dependence of $I_d$ below 150 mK (see Appendix A) conflicts with the thermal broadening explanation, where order of magnitude temperature changes should measurably affect transport. The AC driving current argument does not apply to our DC measurements. The role of vortex interactions in the suppression of the differential resistance peak has been proposed [23], but our peak is absent even in the dilute regime; in addition, large-scale numerical calculations of interacting vortices [8] found a peak in the differential resistance regardless of interaction strength.

Our data supports another explanation. As shown in Fig. 2 (d), the measured linear flux flow region has a nonzero intercept in $I$ and is offset from the simulated curve. Since the intercept occurs near $I_d$, the measured *I-V* curves can smoothly approach flux flow without an inflection point. The lack of an inflection point can generally be attributed to additional dissipation in the system, suggesting that modifications to the dissipation term are necessary to properly model the system.

### III. Simulations

We investigate the dynamics of our system using a phenomenological model built around the Langevin equation, where $N$ vortices are treated as classical objects that propagate under externally applied forces [8, 9]. The classical treatment is valid due to the low resistance of the system, which is overdamped and has suppressed quantum tunneling of vortices. We are also interested in the low vortex-density limit and can expect the vortex motion to occur roughly in a straight line. The dynamics of this system can thus be described using the one-dimensional Langevin equation [24]

$$m\ddot{x}_i(t) = -\frac{\partial V(x_i(t))}{\partial x_i} - \int_0^t \chi(t-\tau)\dot{x}_i(\tau)d\tau + \epsilon_i(t) + \sum_{j=1}^{N} U\left(\frac{x_i(t)-x_j(t)}{L_{int}}\right). \quad (1)$$

In this expression, $m$ controls the inertia of the vortices, $x_i(t)$ is the position of the $i$-th vortex at time $t$, $V(x)$ is the effective potential felt by each vortex, $\chi(t)$ encodes the dissipative interactions between the vortex and the local environment, $\epsilon_i$ is a stochastic force simulating thermal fluctuations, and $U$ is a function that models vortex-vortex interactions. The measured voltage in this system arises mainly from the motion of vortices that travel from one edge of the array to the other and is thus proportional to the average velocity $v = \frac{1}{N}\sum_i^N \dot{x}_i(t)$ of the $N$ vortices.

The effective potential $V(x)$ is approximated by

$$V(x) = V_p \cos\left(2\pi \frac{x}{a}\right) + V_{edge}(x) - J\varphi_0 x \qquad (2)$$

This potential models three properties of the system: a periodic potential of lattice constant $a$ equal to the distance between islands; an energy potential barrier created at the edge of the system due to Meissner currents; and a linear potential that produces a Lorentz force due to the applied current density $J = i/a$. In this expression for $V(x)$, $V_p$ is a parameter representing the strength of the periodic potential and the explicit form of edge potential $V_{edge}(x)$ is given in Appendix B. As a vortex moves through the periodic potential, it slows when crossing potential peaks, lowering the average velocity and measured voltage. The mass term, $m$, suppresses these $\dot{x}(t)$ oscillations. Increasing mass favors a sharp transition from pinned to $V \propto I$ behavior as well as hysteresis. Since we do not observe a sharp transition or hysteresis, $m$ can be assumed to be negligible and is set to zero. This overdamped treatment is consistent with the low resistance of our system.

The dissipation function is commonly written as $\chi(t) = \eta_1 \delta(t)$, which assumes that energy loss occurs due to instantaneous interactions with the environment. This term leads to $V \propto I$ at high currents, where vortex velocity is given by $v = J\varphi_0/\eta_1$. In the low mass and low temperature limit, the current-voltage relationship takes the form $V \propto \sqrt{I^2 - I_d^2}$ (massless particles are greatly slowed when crossing the peaks of the periodic potential when $I \sim I_d$) and there is a differential resistance peak at $I = I_d$. The temperature dependence of $V$ can be solved analytically [12] or simulated by adding a stochastic force, with the results shown in Fig. 3(a). This model converges to $V \propto I$ at large currents, regardless of temperature; this can be contrasted to the experimental data, which shows a non-zero $I$ intercept.

A more general description of dissipation is necessary to explain the flux flow behavior of our experiment. Our model differs from previous treatments in its more general description of dissipation, allowing the inclusion of history dependent dissipative forces in the function $\chi(t)$. An example of a history dependent force is one having a response to a motion event that drops off exponentially with time after that event. Adding this to the dissipative force function leads to $\chi(t) = \eta_1\delta(t) + \eta_2\tau_\beta^{-1}e^{-t/\tau_\beta}$, where $\eta_{1,2}$ are free parameters and $\tau_\beta$ is the timescale of the dissipative force. The effects of a history dependent dissipative force on an overdamped particle are shown in Fig. 3(b), where $\eta_2 = 10\eta_1$ and $\tau_\beta$ is given in terms of $\tau_a = \frac{a^2(\eta_2+\eta_1)}{2\pi V_p}$, which corresponds to the time taken by the large mass particle modeled in Fig. 3(b) to move across one period of the potential at a current infinitesimally greater than $I_d$. A $\tau_\beta$ much shorter than the time taken to cross one period of the potential yields the same behavior as the purely instantaneous dissipative response, but longer $\tau_\beta$ enhances $\dot{x}(t)$ oscillations, leading to very different I-V behavior. When $\tau_\beta$ is much longer than the period crossing time, the dynamic region is highly linear with an I intercept of $I_d$, similar behavior to what we observe in our experiment.

The current-voltage relationship is not strongly dependent on the form of the history dependent component of $\chi(t)$. $\chi(t) = \eta_1\delta(t) + \eta_2 t_c^{-1}\theta(t_c - t)$ achieves similar behavior in the large $t_c$ limit and is less computationally intensive than the exponential expression. The parameters $\eta_2 = 0.4\eta_1$ and $t_c \sim 14\tau_a$ can be used to place the system in the long timescale dissipative force regime, removing the differential resistance peak. Changes in $I_d$ associated with low field Meissner currents and higher field vortex-vortex interactions are achieved by adding an edge barrier and a stochastic force as discussed in Appendix B. Excellent qualitative agreement between theory and experiment can be observed in simulated I-V [Fig. 3(c)] and differential resistance [Fig. 3(d)] plots. This suggests that history dependent dissipation could have a significant contribution to vortex dynamics in overdamped SNS arrays. While this mechanism has previously been considered to study a continuum theory of the plastic flow of vortices [23], the connection to the absence of a peak in the differential resistance was not discussed.

Although the microscopic sources of energy loss are not completely understood [1], one can roughly think of energy dissipation as due to quasiparticles interacting with normal electrons inside vortex cores; the quasiparticles may get excited from impurities in the superfluid, or could leak out of the vortex cores when a current is applied [25]. Memory effects in our system could arise because of a delayed time-scale for the healing of the superfluid density along the path traversed by the vortices as they move through the system. The trail left behind by the vortices would then contribute to the dissipation measured in the experiment.

## IV. Summary

In conclusion, we study vortex motion in SNS arrays, focusing on the dilute vortex filling regime. We find that our observed current voltage relationships are poorly fit by existing models of vortex motion. Instead, our results are consistent with the presence of a history dependent dissipative force in a system of overdamped particles in periodic array. This provides an explanation for deviations from predicted behavior commonly observed in SNS arrays.

## Acknowledgments

The authors wish to thank S. Gopalakrishnan for useful discussions. This work was supported by the DOE Basic Energy Sciences under DE-SC0012649. IMS acknowledges support from the Sloan Foundation. This research was carried out in part in the Center for Microanalysis of Materials, UIUC.

## Appendix A: Temperature dependence of *I-V* measurements

The temperature dependence of 390 nm edge to edge spaced arrays is discussed here in greater detail. *I-V* measurements were performed at fixed temperature intervals at low frustrations. As shown in Fig. 4 (a), we observe the suppression of $I_d$ and $I_c$ with increasing temperatures, but the $I_d$ and $I_c$ curves flatten at fixed values below 150 mK and 200 mK respectively. A device with 440 nm edge to edge spaced islands, situated on the same chip and measured during the same run as the 390 nm device, shows strong temperature dependence at lower temperatures than the 390 nm array. This suggests that the weak temperature dependence of the 390 nm sample below 150 mK is not due to heating or electron temperature issues in our measurement apparatus, but that the array is in a regime not dominated by thermal activation.

The temperature dependence of 390 nm islands at f = 0.007, shown in Fig. 4 (b), is qualitatively consistent with the delayed dissipative force model presented in this paper, with increasing temperatures providing a stochastic force that effectively weakens vortex pinning and yielding parallel *I-V* curves in the flux flow regime. It is inconsistent with finite temperature RCSJ model predictions that the *I-V* curves should converge at the $I_d$ [26].

## Appendix B: Vortex dynamics at low frustrations

In this section, we provide some details of the numerical simulations. For completeness, we recall that the vortex dynamics is described by the generalized Langevin equations of motion given in Eq. (1). In the main text we have already described the main approximations that are implemented in the numerical simulation. In what follows, we will clarify additional technical details.

### 1. Effective vortex potential

We approximate the effective potential $V(x)$ by

$$V(x) = -J\varphi_0 x + V_{edge}(x) + V_p \cos\left(2\pi \frac{x}{a}\right),$$

where

$$V_{edge}(x) = -\left[\frac{i_e(f)\ell}{1+\left(\frac{x}{\ell}\right)^2} + \frac{i_e(f)\ell}{1+\left(\frac{x-L}{\ell}\right)^2}\right].$$

The parameter $\ell$ is the length scale over which the edge potential varies, and $i_e(f)$ encodes the depth of the edge potential which we will approximate to decrease linearly with frustration $i_e(f) = (1-f/f_c)\theta(f_c - f)i_0$. The step function is introduced to describe the fact that, at a characteristic frustration, vortices that are formed in the bulk will dominate the voltage that is detected, which represents the end of region A and the beginning of region B in Fig 5(a).

### 2. Dissipative dynamics

As we explained in the main text, the function $\chi(t)$ encodes the degree to which the dissipation of the vortices is correlated in time. This correlation usually arises because generically it takes some time for the bath to react to the presence of the moving vortices. To simplify the calculation, we divided this response function into two pieces

$$\chi(t) = \chi_1(t) + \chi_2(t).$$

The first contribution corresponds to the environment responding instantaneously to the vortices, which can be modeled as

$$\chi_1(t) = \eta_1 \delta(t).$$

If this were the only contribution to the dissipation, there would be only history-independent dissipation. In this case, the solution to the Langevin equation has been computed analytically [12], and takes the form

$$\tilde{V}(\tilde{I},\tilde{T}) = \frac{\tilde{T}\left[1-e^{-\pi(1/\tilde{T})}\right]}{\int_0^{2\pi} e^{-u(1/\tilde{T})} I_0\left(\frac{2}{\tilde{T}}\sin\frac{u}{2}\right) du} \qquad (B1)$$

where $I_0(x)$ is the modified Bessel function of zero order, and the variables $\tilde{T} = \left(\frac{k_B T}{V_p}\right)$, $\tilde{V} = \left(\frac{L/a}{4 r_n I_d}\right) V$, $\tilde{I} = (I_d)^{-1} I$ are the normalized temperature, voltage and current, respectively. Here $L$ is the linear size of the Josephson array. Upon taking a first derivative with respect to the normalized current, one obtains the differential resistance curves shown in Fig. 3(a) for a set of temperatures. It can be seen that a peak in the differential resistance persists even when the temperature is increased. As is discussed in the main text, this runs contrary to the experimental measurements, thus justifying the use of a more general form of dissipation.

The second contribution to the response function, $\chi_2(t)$, encodes the lag in the reaction of the environment to the interaction with the moving vortices, which we approximate by a flat function that correlates times on the scale $t_c$

$$\chi_2(t) = \eta_2 t_c^{-1} \theta(t_c - t).$$

In order to be consistent, we need to satisfy the fluctuation-dissipation relation

$$\langle \tilde{\epsilon}(t_1) \tilde{\epsilon}(t_2) \rangle = F \chi(t_1 - t_2) = F\left[\eta_1 \delta(t_1 - t_2) + \eta_2 t_c^{-1} \theta(t_c - |t_1 - t_2|)\right],$$

where $F$ parameterizes the strength of the stochastic forces acting on the vortices. We need to obtain a distribution $\tilde{\epsilon}(t)$ such that this correlation function is satisfied. First, we note that in the numerical simulation we need to discretize the time domain in $N_s$ steps of size $\delta t$. Because of this, we can approximate the response function as

$$\chi(t_1 - t_2) \approx \chi_{n_1, n_2} = \frac{\eta_1}{\delta t} \delta_{n_1, n_2} + \frac{\eta_2}{t_c} \sum_{m=1}^{\lfloor t_c/\delta t \rfloor} \delta_{|n_1 - n_2|, m},$$

where $t_j = n_j \delta t$. We can thus obtain the required stochastic force $\tilde{\epsilon}(t)$ by performing the so-called Cholesky decomposition of the square matrix $\chi_{n_1, n_2} = K^T K$. The stochastic force then takes the form $\tilde{\epsilon} = K\xi$, where $\xi$ is a list of $N_s$ random variables satisfying the correlation matrix $\langle \xi_{n_1} \xi_{n_2} \rangle = \delta_{n_1, n_2}$.

For small matrices, it can be checked analytically that, to lowest order in $\frac{\eta_2}{\eta_1}\frac{\delta t}{t_c}$, the expression for the correlated random force is given by

$$[\tilde{\epsilon}]_n \approx \sqrt{\frac{F\eta_1}{\delta t}}\left[\xi_n + \frac{\eta_2}{\eta_1}\frac{\delta t}{t_c}\sum_{m=1}^{\lfloor t_c/\delta t \rfloor}\xi_{n+m}\right].$$

In this expression, the correlations of the random numbers are encoded in the second term in brackets. One can see that this contribution is suppressed by the fact that it is proportional to the sum of uncorrelated random numbers that have zero mean. Hence, we can approximate $[\tilde{\epsilon}]_n \approx \sqrt{\frac{F\eta_1}{\delta t}}\xi_n$.

For larger matrices, one can also confirm numerically that most realizations of the random variables abide closely by this approximate expression. Thus, we conjecture that this form of the random variable $[\tilde{\epsilon}]_n$ is a good approximation of the correlation matrix. In view of the fact that the contribution from correlations is clearly suppressed in the expression for $[\tilde{\epsilon}]_n$, we have approximated the $[\tilde{\epsilon}]_n$ to be uncorrelated in the numerical simulations.

3. **Explicit equation of motion and parameters used**

In what follows, we will use the following units: $a=1$ (length), $I_d = \frac{2\pi V_p}{\varphi_0 a} = 1$ (barrier), $\eta_1 = 1$ (dissipation strength), $t_0 = \frac{\eta_1 a}{\varphi_0 I_d} = 1$ (time). Implementing these approximations and rearranging terms, we obtain the final expression:

$$\dot{x}_i(t) = J - \sin(2\pi x_i(t)) + i_e(f)\left(\frac{2x_i(t)/\ell}{\left[1+\left(\frac{x_i(t)}{\ell}\right)^2\right]^2} + \frac{2(x_i(t)-L)/\ell}{\left[1+\left(\frac{x_i(t)-L}{\ell}\right)^2\right]^2}\right) - \eta_2\frac{(x_i(t)-x_i(t-t_c))}{t_c} + \tilde{\epsilon}_i(t).$$

We numerically propagated in time this simplified equation for $N=500$ realizations of the stochastic force $\tilde{\epsilon}_i(t)$. More specifically, we computed the average time $t_{ave}$ it took the vortices to traverse the system from one edge to the other, from which the average velocity is $v_{ave} = L/t_{ave}$. As we mentioned in the main text, the average of the resulting velocities of the vortices provides a measure of the voltage measured in the experiment $V = fv_{ave} = fL/t_{ave}$, where we multiplied by the frustration in order to take into account the number of vortices actually in the system. For the calculations presented in the main text, we used the parameters $L = 60a$, $t_c = 20\ t_0$, $\ell = 10\ a$, $i_0 = 30 I_b$, $\eta_2 = 0.4\eta_1$ and $F = 0.05$. The frustration spans the three regions of Fig. 5a. To better illustrate the effect of history-dependent dissipation, in Fig. 5(b) we show how the peak gets suppressed as the parameter $\eta_2 t_c^{-1}$ is increased. We caution that there could be other values of the parameters that lead to similar results and, in particular, exhibit the same suppression of the peak in the differential conductance.

**Figures.**

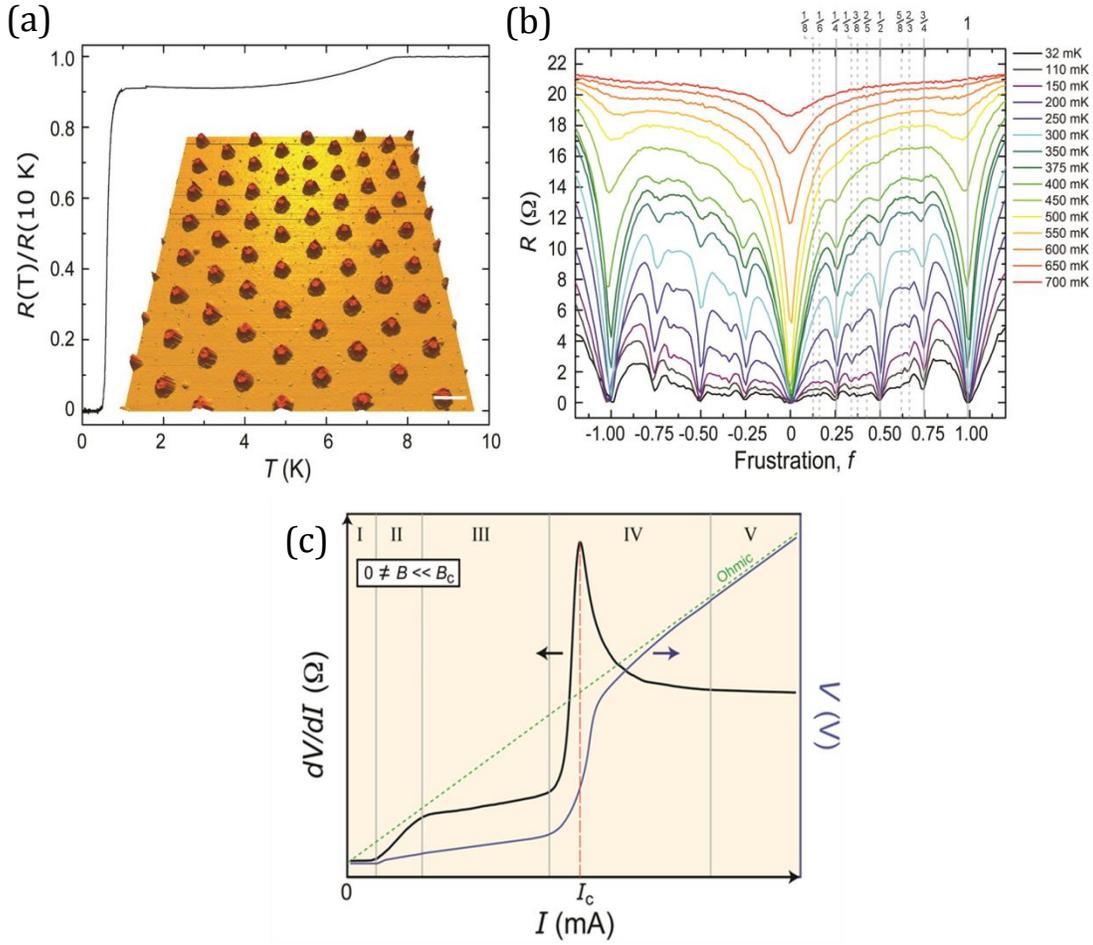

FIG. 1. Transport measurements on an array of Nb islands on a 10 nm thick Au layer. The island spacing is $d = 540$ nm and island height 125 nm. (a) Resistance $R$ (normalized to values at 10 K) versus temperature $T$, showing a two-step transition to superconductivity at zero-field. Inset is an atomic force microscopy device image, with scale bar of 500 nm. (b) Magnetoresistance $R$ vs. frustration $f$ at different temperatures (specified in the legend). Dips are present at certain rational values of frustration. The gray vertical lines specify frustrations at which field-induced vortices are strongly commensurate (solid lines) and weakly commensurate (dashed lines) with the Nb island lattice. (c) As the current increases, vortices are initially pinned in region I, exhibit measurable vortex creep in region II, and freely flow at a terminal

velocity in region III. Above a critical junction current, $I_c$, the array transitions into Ohmic behavior in regions IV and V.

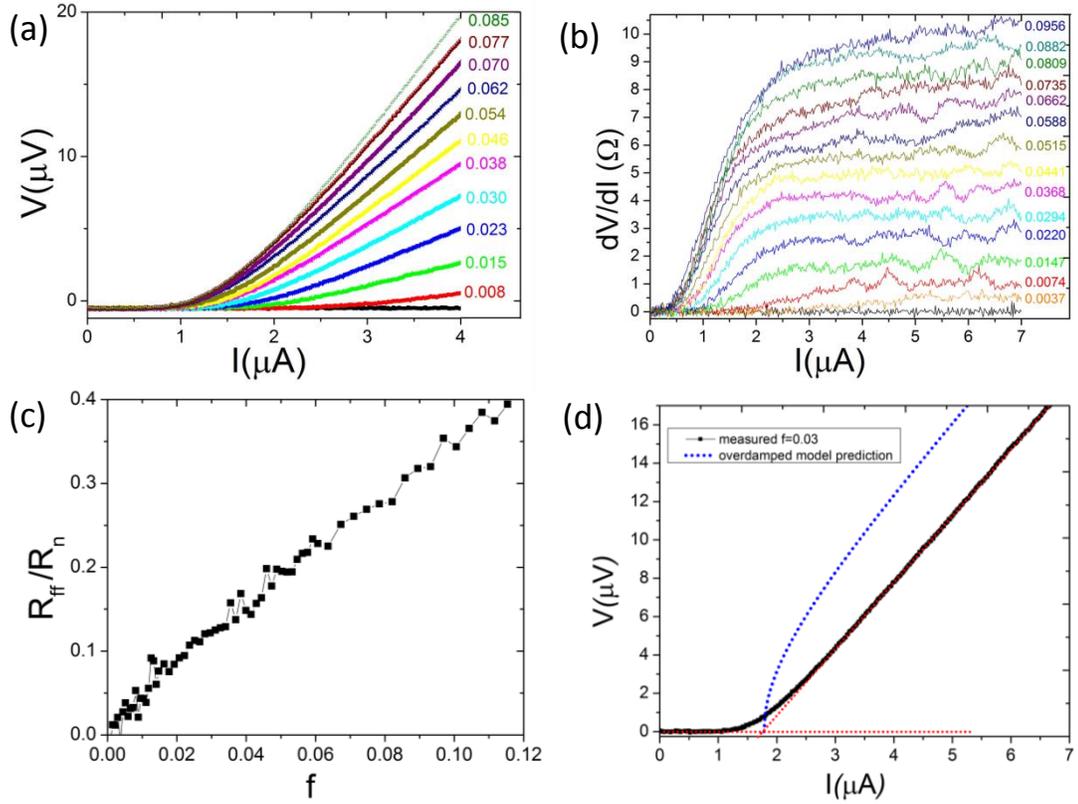

FIG. 2. Current-induced vortex de-pinning for 390 nm spaced islands at $T = 17$ mK. (a) *I-V* measurements performed using a swept DC current bias for $I < I_c$ in different magnetic fields. Adjacent numbers indicate the frustration associated with each curve; $f = 1$ corresponds to magnetic field $B = 115$ gauss. (b) Differential resistance, $dV/dI$, extracted from *I-V* measurements in (a). (c) The flux flow resistance $R_{ff}$ vs $f$ extracted from *I-V* curves. $R_{ff}$ is normalized to the normal state resistance $R_n$. (d) Measured *I-V* at f = 0.03 (black) compared with the prediction of the overdamped vortex model (dashed blue). A linear fit is performed on the superconducting and flux flow regions of the measured curve (red dotted lines). These intersect at a nonzero *I* intercept.

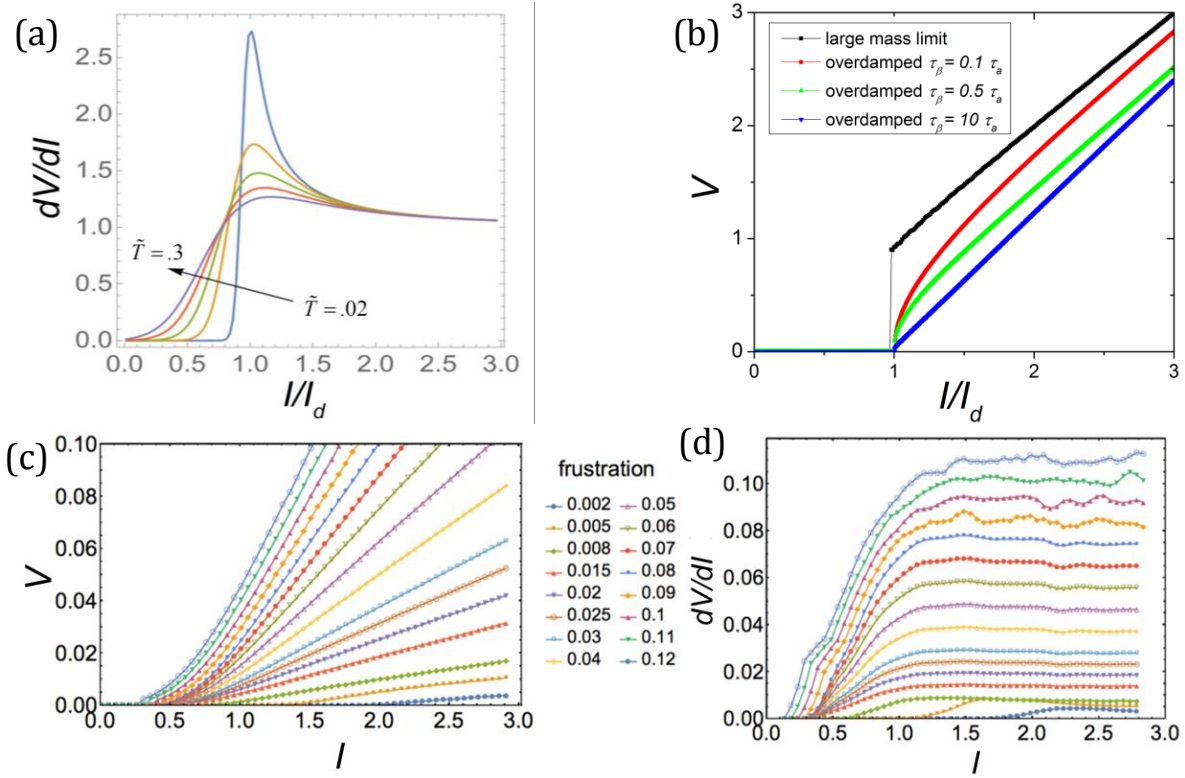

FIG. 3. Simulated Vortex Dynamics (a) The temperature dependence of an overdamped non-interacting vortex system with instantaneous dissipation (Eq. B1). The arrow denotes the direction of increase of the temperature, which is given in terms of normalized temperature, $\tilde{T} = k_B T / V_p$. The peak survives even when the temperature is large enough to cause considerable thermal creep near $I = 0$, which we do not observe in our experimental measurements. (b) Simulated $I$-$V$ behavior showing the effects of the mass term and the timescale of the dissipative force, $\tau_\beta$, in a purely periodic potential. Three curves show the predictions for the low mass limit with three different dissipative force time scales. $\tau_\beta = 0.1\, \tau_a$ is indistinguishable from the instantaneous dissipative force time constant case, but much longer time scales result in a linear region with an $I$ intercept near $I_d$. (c) Simulated voltage vs current and (d) differential resistance vs current for a generalized Langevin equation with time dependent dissipation where $\tau_\beta \sim 14\, \tau_a$.

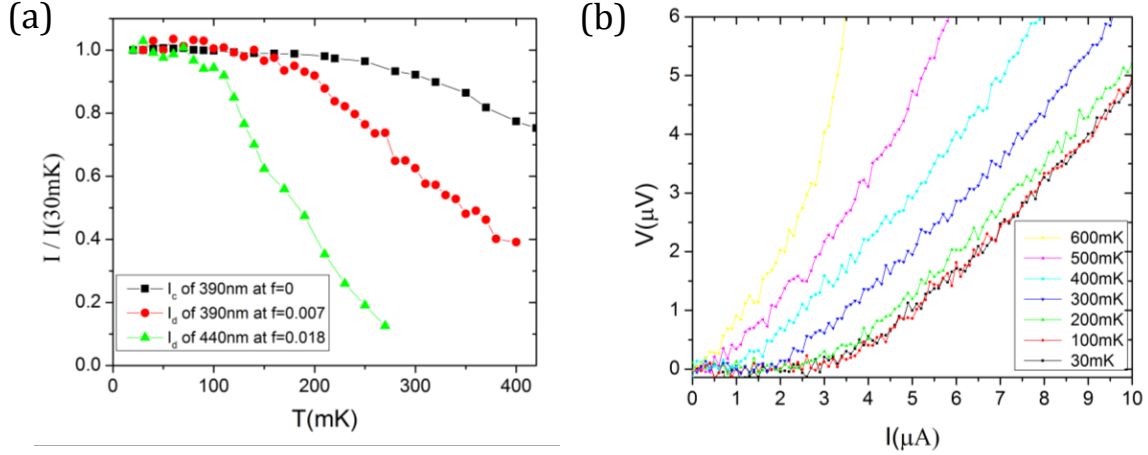

FIG. 4. (a) $I_d$ and $I_c$ temperature dependence for 390 nm spaced islands at low field values as well as $I_c$ temperature dependence for 440 nm islands at $f = 0.018$. (b) *IV* measurements performed in 390 nm edge to edge spaced arrays at $f = 0.007$ at various temperatures.

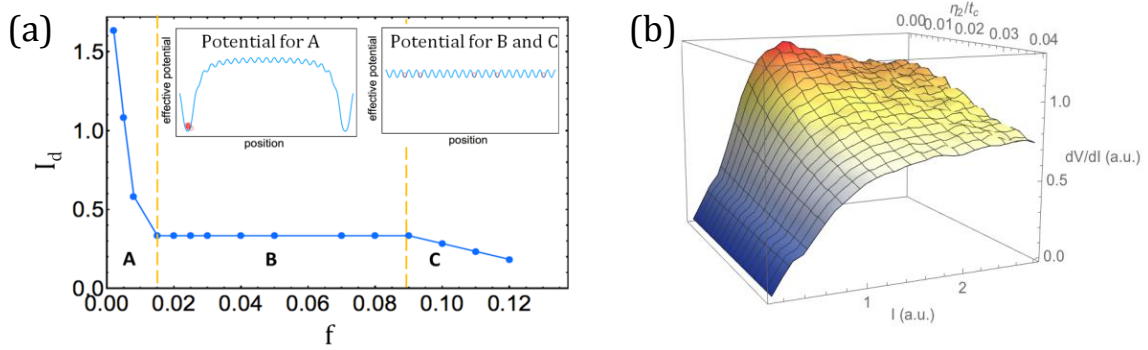

FIG. 5. (a) A schematic showing the potentials used in Fig. 3(c)(d). Region A is heavily influenced by an edge barrier that drops off rapidly with magnetic field. In Region B the edge barrier is negligible and vortices are pinned to the periodic potential from the island lattice, shown in the right inset. Region C simulates vortex-vortex interactions by increasing the stochastic force. (b) Simulated differential resistance as a function of applied current and the parameter $\eta_2 t_c^{-1}$. The remaining parameters are fixed to the same values used for the present work. It can be seen how the history-dependent dissipation suppresses the peak in the differential resistance.